\documentclass[12pt]{article}
\usepackage{amssymb}
\catcode`\@=11
\def\defop#1{\expandafter
\def\csname#1\endcsname{\mathop{\rm#1}\nolimits}}
\defop{Map}  \defop{Mat}   \defop{Gr}   \defop{tr}
\defop{Tr}
\defop{Det}  \defop{str}   \defop{Str}  \defop{sdet}
\defop{Sdet}  
\defop{Cliff}\defop{diag} \defop{sp}    
\defop{Cl}
\let\@@line\|       

\def\|#1>{\mathopen|#1\rangle}

\def\^#1#2#3{\langle#1|#2|#3\rangle}
\def\<#1>{\left\langle#1\right\rangle}
\def\=#1{{\bf#1}}
\def\-#1{{\bar{#1}}}

\def\norm#1{\mathopen\@@line#1\mathclose\@@line}
\def\slash{\@ifnextchar[{\@slash}{\@slash[\z@]}}
\def\@slash[#1]#2{
\setbox\z@\hbox{$#2$}\@tempdima\wd\z@\box\z@%
\@tempdimb#1 \advance\@tempdimb-\@tempdima \kern\@tempdimb
\hbox to\@tempdima{\hss\@makeslash\hss}}
\def\@makeslash{$/$}
\catcode`\@=12
\newcommand{\beq}{\begin{equation}}
\newcommand{\eeq}{\end{equation}}
\newcommand{\be}{\begin{eqnarray}}
\newcommand{\ee}{\end{eqnarray}}
\newcommand{\dd}{{\mathrm d}}
\newcommand{\e}{{\mathrm e}}
\newcommand{\de}{\partial}

\newcommand{\Dslash}{{\slash{D}}}
\newcommand{\ua}{{\underline{\alpha}}}
\newcommand{\ub}{{\underline{\beta}}}

\newcommand{\uab}{{\underline{\alpha\beta}}}
\newcommand{\half}{{1 \over 2}}

\begin{document}
\begin{titlepage}
\begin{flushleft}
       \hfill                      {\tt hep-th/0007234}\\
       \hfill                       SISSA 77/2000/EP\\
\end{flushleft}
\vspace*{2mm}
\begin{center}
{\bf \Large Holographic Renormalization Group}\\
\vspace*{3mm}
{\bf \Large with Fermions and Form Fields}
\vspace*{8mm}

{\large Jussi Kalkkinen\footnote{\tt kalkkine@sissa.it}}
$\qquad$ and $\qquad$
{\large Dario Martelli\footnote{\tt dmartell@sissa.it}}\\
\vspace*{4mm}
{\it SISSA, Via Beirut 2 Trieste 34014, Italy\\}
{\it and\\}
{\it INFN, Sezione di Trieste\\}
\vspace*{5mm}
\end{center}

\begin{abstract}
We find  the Holographic Renormalization Group equations 
for the holographic duals of generic gravity theories 
coupled to form fields and spin-$\frac{1}{2}$ fermions.
Using Hamilton--Jacobi theory we discuss   
the structure of Ward identities, 
anomalies, and the recursive 
equations for determining the divergent terms 
of the generating functional. 
In particular, the Ward identity associated to diffeomorphism invariance
contains an anomalous contribution that, however, 
can be solved either by a suitable counter term or  
by imposing a condition on the boundary fields.
Consistency conditions for the existence of the dual 
arise, if one requires that a Callan--Symanzik type 
equation follows from the Hamiltonian constraint. 
Under mild assumptions we are able to find a class of solutions 
to the constraint equations.
The structure of the fermionic phase space and the constraints are 
treated extensively
for any dimension and signature.
\end{abstract}
\end{titlepage}

\tableofcontents 

\section{Introduction}

In theories with diffeomorphism invariance the Poincar\'e generators are gauged, and the
concept of a point has no invariant meaning. It becomes therefore all but impossible to
define local observables. The only known way around this problem is provided by  
Holography \cite{'tHooft93,Susskind:1995vu}, which means 
(in the strong form) that a theory with diffeomorphism invariance 
should be describable in terms of 
a dual local quantum field theory defined on the boundary of space-time. 
In this article we investigate the conditions that a diffeomorphism invariant
theory of interacting fermions and form fields has to satisfy in order that 
its holographic dual, in the sense defined presently, exists. 

The AdS/CFT conjecture \cite{malda} has provided a very concrete example
of the Holographic Principle, asserting that the physical 
content of a theory of gravity in $d+1$ dimensions should be encoded in a  
quantum conformal field theory living on the boundary of its anti de Sitter
space-time.
The correspondence has been made precise in \cite{GKP,Witten1} 
where it was explained how to compute
field theory observables (Green functions) in terms of AdS integrals.
The central object in this 
correspondence is the generator of connected Green functions
$W$ 
\be
\e^{-W[g,J]}=\int {\cal D}\Phi~ \e^{-S[g,\Phi]-J {\cal O}[\Phi]}
\ee
where $J$ is, on the QFT side, a set of 
external sources coupled to (composite) operators $\cal O$ 
and $g_{ij}$ is a background
metric, which allows to compute the stress-energy tensor. 
The conjecture then states that
\be
W[g,J]=S_\mathrm{SUGRA}[g,J]
\ee
where the on-shell supergravity action is evaluated on an AdS solution, and it is
a functional of initial values $J$ of the fields appearing in it. 
The quantum field theory content is therefore encoded
in a classical field theory for its sources.

This correspondence has been later generalized to non-conformal theories
\cite{Girardello98,Distler98,Nojiri:1999yx,KehagiasHRG,PorratiHRG,BalasubraHRG,Freedman99}. 
There the classical radial equations of motion on the supergravity side
were interpreted as the renormalization group (RG) flow. In this way
supergravity solutions provide information about RG trajectories
connecting different CFT's appearing at the fixed points of the 
flow\footnote{This mechanism was discussed in \cite{Ferretti} in the framework 
of Polyakov's proposal for holographic noncritical strings \cite{thewall,language}.}. 
In this bulk/boundary correspondence the central object is still $W$, which
is now a functional of the sources at a given mass scale. 

In QFT $W$ obeys a set of Ward identities, which ensure that classical 
symmetries are preserved at the quantum level and, consequently, constrain 
the counter terms that can be used in renormalization. For instance, 
the Ward identity of (broken) scale invariance
is the Callan--Symanzik equation, which is a particular form of RG equation.
Much of the information contained in this structure can be traced back to 
divergences.  Regularization of these infinities introduces a scale, and 
the requirement that the renormalized quantities are independent of the cutoff
produces the RG equation. In this article, we will discuss  how 
the holographic renormalization group (HRG) manages, generically, to produce these 
features, but also how, and why, it may sometimes seem to fail if the theory includes 
interacting fermions.

\vspace{1cm}

As Holography relates theories in $d$ and $d+1$ dimensions it is useful
to resort to Hamiltonian formalism, where the transverse ``time'' coordinate
plays a distinguished role. As we are particularly interested in the 
on-shell action it is rather natural, as proposed in 
\cite{dBVV}, to use Hamilton--Jacobi theory. There are also some 
conceptual reasons for doing so: Most importantly, 
on the QFT side, one eventually needs a formulation 
in terms of first order equations, as they can be interpreted as RG flow 
equations. Also, on the gravity side, it is natural to expect
the Hamiltonian formulation to arise \cite{spinors3}, as one considers  
classical gravity as a WKB limit of quantum gravity (String/M-Theory). 

In the Hamiltonian formalism one chooses a coordinate $t$, with 
tangent vector field $e_0$, to parameterize the evolution of the 
initial value hypersurface, the ``boundary''. It is sufficient to characterize this 
hypersurface by giving its normal direction $n$ with norm $n\cdot n=\pm 1$. 
Depending on this sign, denoted  $\eta$, one is considering either time-like or space-like boundaries. 
This will produce sign changes with respect to usual Minkowskian Hamiltonian
theory, and we will most efficiently keep track of them by leaving  $\eta$ 
unspecified. Otherwise we follow the standard Hamiltonian
reduction. 
Writing the full metric 
\beq
\dd s^2=(\eta N^2 + N_i N^i)\dd t^2 +2 N_i \dd t\dd x^i 
+ g_{ij}\dd x^i \dd x^j 
\label{ADMmetric}
\eeq
the canonical (ADM) gravitational action coupled to matter has the form 
\be
S = \int \dd t \left\{ p~\dot{q} -\int\dd^d x \; N {\cal H}_{\perp}+ 
N^i {\cal H}_i + \Theta^A {\cal G}_A \right\}
\ee 
where $p\dot{q}$ is a shorthand for the kinetic term of the dynamical degrees of
freedom, $N$ and $N^i$ are the lapse and shift functions, which together
with $\Theta^A$ are the Lagrange multipliers for the 
constraints\footnote{These are first class constraints. We will
consider also second class constraints. They appear in the presence of fermionic 
fields that enter the action linearly.}
\be
{\cal H}_\perp \approx {\cal H}_i \approx {\cal G}_A\approx 0~.
\ee   
These constraints generate, in the bulk theory, orthogonal
deformations and diffeomorphisms on the initial value hypersurface and 
possible additional local symmetries, such as gauge, local Lorentz or 
supersymmetries. 

In the Hamilton--Jacobi theory one makes a canonical transformation such that  
the new phase space coordinates are constants of motion. In gravitational theories
the generating functional $F$ of the canonical transformation must satisfy
simultaneously the constraints 
\be
{\cal H}_\perp (q,\frac{\delta F}{\delta q}) & = &0 \label{CS}\\
{\cal H}_i (q,\frac{\delta F}{\delta q}) & = & 0 \label{DiffWI}\\
{\cal G}_A (q,\frac{\delta F}{\delta q}) & = & 0~.\label{WI} 
\ee
The Hamilton principal function $F[q]$ 
is actually just the classical action evaluated at some given time $t$ 
for fixed boundary values $q(t)$. The momenta can be calculated from 
\be
p=\frac{\delta F[q]}{\delta q}~. \label{pqF}
\ee
This gives us the action; if we want to calculate the full equations of motion,
we have to consider also 
\beq
\dot{q}=\{q,H\}_{\scriptscriptstyle \mathrm{DB}}~,\label{flow}  
\eeq 
where $\{\cdot,\cdot\}_{\scriptscriptstyle \mathrm{DB}}$ denote
the Dirac brackets.
The holographic correspondence allows us to give a dual field theory 
interpretation to the equations above. 

Halving the degree of the equations of motion  
by using Hamilton--Jacobi equations for $F$, in spite of giving a rather appealing
form for application to holography, boils down to having a high degree
of arbitrariness in $F$. Equations (\ref{CS}), (\ref{DiffWI}), and (\ref{WI}) 
have in fact a huge number of solutions, many of them unacceptable on
physical grounds.
This was discussed\footnote{There are some comments in this direction in
\cite{modeling}, as well.} in \cite{Martelli}, for a scalar field 
coupled to gravity.
One should keep in mind also the fact that not every gravity solution   
corresponds to deformations of a CFT by adding
an operator to the fixed point Lagrangian, as was pointed out in
\cite{Balasubraplusminus,KWplusminus}. 
Rather, some of them may correspond to an altogether different vacuum of 
the same theory, where an operator has acquired a nontrivial vacuum expectation 
value\footnote{For a clear exposition of this point
see for instance \cite{vevdef}. This was illustrated explicitly in 
\cite{Martelli} in connection to the Hamilton--Jacobi formalism.}.  
In general, only a subset of solutions will admit a direct physical 
interpretation \cite{gubser}.
To ensure a physically acceptable result one specifies 
the asymptotic (near the AdS boundary) behaviour of the solutions.
If one wishes to solve the equations for a coupled system, 
with a dynamical metric or interacting sources, in a perturbative fashion,
the back-reaction will involve the flow equations, which determine the
scaling behaviour of the fields, modifying then the general ansatz for $F$ ---
even for its local part, which is the only one computable in general.

\vspace{1cm}

Finally, let us briefly comment on how the QFT divergences arise on
the gravity side. It is well known from explicit calculations that
the Einstein--Hilbert action (with the cosmological constant and the 
Gibbons--Hawking term) is divergent when evaluated on asymptotically
AdS solutions. 
For instance in the coordinate system 
\be
\dd s^2 = \frac{1}{t^2}(\dd t^2 + g_{ij}^R\dd x^i \dd x^j)
\label{rsquare}
\ee
the determinant gives $\sqrt{g}=t^{-d}\sqrt{g^R}$, and 
$R_{ij}= t^2 R_{ij}^R$, so that divergences
arise in the limit $t \to 0$ (asymptotically AdS boundary) for terms containing
up to $[\frac{d-1}{2}]$ $\mathrm{Ricci}$'s. 
The existence of such solutions in this context is
a physical requirement: It means that the RG has fixed points.
In general, including additional terms in the bulk action will introduce 
further divergences.

In this article we shall consider the fermions and form fields with some 
apriorously fixed anomalous dimensions at the UV conformal point, and
eventually consider  how  general the results are.
This is tantamount to restricting the analysis to a particular subset of
operators used to flow from the UV CFT. 
In even boundary dimension there can be additional logarithmically 
divergent terms that, in the Hamilton--Jacobi context, can arise only from 
{\it nonlocal} contributions.
One should isolate the terms in $F$ that are divergent
in the CFT limit: These terms will give rise to
the beta functions, while the rest will give the renormalized  generating
functional.

It turns out that -- in analogy with what is achieved by QFT Ward identities -- 
these local divergent terms, which we will call $S_\mathrm{div}$,
are fixed by the constraints in the classical theory, and 
can indeed be determined recursively order by order according to their 
degree of divergence. 
This recursive procedure can be expressed as ``descent equations'' 
and is computationally 
equivalent to the counter term generating algorithm proposed in
\cite{Kraus1,Ho} in the context of pure AdS gravity\footnote{In that
context form fields were considered in \cite{marika}.}.
Performing this calculation one sees that (\ref{DiffWI}) and 
(\ref{WI}) will produce Ward identities 
(diffeomorphism and gauge symmetries), while
(\ref{CS}) provides a recursive equation for 
$S_\mathrm{div}$, formally very similar to the perturbative expansion of
the master equation, that in addition 
will give rise to a Callan--Symanzik-type equation for $W$. 

One should naturally try to chose a renormalization procedure 
that preserves the symmetries. A failure to fulfill this requirement 
would give rise
to an anomaly. Anomalies in the boundary theory may  arise from
bulk contributions to the Hamiltonian constraints. 
This may happen because the Hamiltonian ``time''-slicing  
might not preserve local bulk symmetries after gauge fixing. An example is
provided by the holographic chiral anomalies discussed in \cite{Witten1}, and in
\cite{holoWard} in the Hamilton--Jacobi framework.

\vspace{1cm}

After the work of \cite{dBVV} there has been a number of 
papers discussing different
aspects of the HRG using the Hamilton--Jacobi formalism for gravity. Most
of them, however, elaborate gravity coupled to scalar fields 
\cite{Nojiri:2000kh,otherholo1,otherholo2,otherholo3,otherholo4,Martelli,Akhoury:2000zj,Skenderis1,Cfunction},
where the situation is by now well understood. In this article
we shall extend this analysis to essentially more general situations. We consider
a general setup in which spin-$\half$ fermions and form fields interact with
gravity. In particular the treatment
of spinors is rather subtle. These fields  are linear in momenta,
couple derivatively to the metric and are in general endowed with a complex
structure. It was already noticed in previous literature \cite{spinors1,spinors2,spinors3,spinors4} 
on the AdS/CFT 
correspondence that fermion fields split naturally in momentum and coordinate parts,
and that the latter part are the relevant sources for the boundary operators.
The split boils down, in general, to imposing some  Lorentz-invariant 
condition on the fermions, such as chirality or reality conditions. 
In the following we systematize these observations in the context of Holography,
using the Hamilton--Jacobi approach.
The Ward identities resulting in this case from the gravitational constraints 
generalize the structures that arise from the scalar and Yang--Mills sectors,  
and display some intriguing novel features, as well.

The plan of the paper is as follows. In Section 2 we use the 
pure gravity system to illustrate how to implement and solve the descent
equations that follow from the ${\cal H}_\perp$ constraint. 
In the process we will 
present a novel way of calculating  the holographic  Weyl anomaly of \cite{HS}\footnote{A similar 
approach has recently appeared in \cite{otherholo4}.}.
We make then general comments on how Ward identities and holographic
anomalies arise. In Section 3 we briefly review the structure of the holographic
Callan--Symanzik equation and notice that there exists a distinguished coordinate
system in which the flow equations take  the familiar form involving beta 
functions.
Section 4 is devoted to explaining the Hamiltonian treatment of spin-$\half$
fermions and form fields. Sections \ref{bracket}, \ref{form}, and \ref{locexp} 
contain our main results. In Section  \ref{bracket} 
we present the bracket structure induced by the ${\cal H}_\perp$
constraint, 
while in Section  \ref{form}  we discuss the other Ward identities.
In Section \ref{locexp} we solve, under certain (mild) assumptions, 
the bracket equations. We find agreement with results concerning the pure
gravity sector and some interesting constraints in the form field and fermion
sectors. The last section
contains conclusions and future perspectives. In the appendices we have
relegated some useful formulae and notation, and an extensive
analysis of the  fermionic phase space.

\section{Renormalization and Anomalies}
\label{RA}

In order to illustrate the procedure, we consider in this section 
pure gravity with a cosmological constant. We start by showing how to get
a solution for local terms of the generating functional $F$. As a by-product we shall get an 
alternative derivation of the holographic Weyl anomaly computed in \cite{HS}.
We then set the ground to discuss Ward identities and anomalies.

Equation (\ref{CS}) in this case reads
\beq
   \frac{-\eta\kappa^2}{\sqrt{\hat{g}}} \left( g_{ik}g_{jl}-
  \frac{1}{d-1} g_{ij}g_{kl} \right) \frac{\delta F}
  {\delta g_{ij}} \frac{\delta F}{\delta g_{kl}}=
  \sqrt{\hat{g}}\left(\frac{1}{\kappa^2}R-\Lambda \right)
  \label{HJpuregrav}
\eeq
and, following the notation introduced in \cite{VV}, we rewrite it 
as\footnote{The subscript counts the number of derivatives.}
\beq
(F,F)={\cal L}_{0}+{\cal L}_{2}~.\label{CS1}
\eeq
We then expand $S_\mathrm{div}$ according to the degree of divergence of
each local term. For the case of pure gravity this is equivalent to a
derivative expansion, as is easily seen by a change of
coordinates $x_i \to t^{-1} x_i$.

The local terms that diverge at the UV fixed point are those
that contain up to $d-1$ derivatives; logarithmically divergent terms may 
arise in the nonlocal part of $W$. 
However, the most divergent ones come always multiplying a scale invariant
term. This is typical of the structure that arises in effective actions
that produce a conformal anomaly \cite{Deser-anomaly}.

Let us set   $F=S_\mathrm{div}+W$ in (\ref{HJpuregrav}) and expand 
order by order in $t$. In global scalings the metric behaves as 
$g_{ij}\to t^{-2} g_{ij}$. Given that 
\be
(\cdot,\cdot) & \to & t^{d}(\cdot,\cdot)\\
S_n & \to & t^{-d+n}S_n\\
W & \to & W - 2\log t~ S_d~,
\ee
and $(S_0,S_d)=0$, the following descent equations have to be satisfied 
\be
{\cal L}_{0}   &=&    (S_0,S_0) \nonumber \\
{\cal L}_{2}    &=&    2(S_0,S_2) \nonumber \\
0 &=& 2(S_0,S_4)+(S_2,S_2) \label{fourth}\\
&\vdots& \nonumber \\
-2(S_0,W)  &=&  2(S_2,S_{d-2}) + \cdots 
+ 2(S_{\frac{d}{2}-2},S_{\frac{d}{2}+2}) + (S_{\frac{d}{2}},S_{\frac{d}{2}})~.\nonumber
\ee
The zeroth order equation fixes the relationship between the cosmological constants in the 
bulk and on the boundary, see Eq. (\ref{LL}).
The $n$'th order equation is an equation for $S_n$ in terms of 
$S_2,\dots,S_{n-2}$. In particular, 
the last equation\footnote{In even boundary dimension. In odd dimensions
it fixes also higher order terms, and nonlocal logarithmic terms do not arise,
as is well known.} then gives the trace of the bare 
stress-energy tensor.  
This is because the lowest order term $S_0$ acts through the bracket
$(\cdot,\cdot)$ 
essentially as a scale transformation. The trace anomaly at the conformal
points, where the beta functions vanish and couplings go to their fixed
point values $J^*$, is given by
\be
\<T> & = &  \frac{2\eta (d-1)}{\kappa^2 \hat{\Lambda}}
\lim_{J\to J^*}\frac{1}{\sqrt{\hat{g}}}~ (S_0,W)~.
\ee 
In the UV limit this expression remains finite and reproduces the holographic 
Weyl anomaly of \cite{HS}. The results for $d=2$ and $d=4$ are\footnote{With
$\eta=1$, see equations (\ref{LL}) and (\ref{v1}).}
\be
\<T> & = & \frac{1}{\kappa^4 \hat{\Lambda}}~R \label{dtw}\\
\<T> & = & \frac{k_1^2}{\hat{\Lambda}}~\left( 3R^{ij}R_{ij}-R^2\right)~.
\ee 

The only Ward identity that one has to check in this case is diffeomorphism
invariance, Eq.~(\ref{DiffWI}). 
It is trivially satisfied for the finite (bare) stress-energy tensor,
because the counter terms are generally covariant in the boundary metric 
\be
\nabla_i \frac{\delta W}{\delta g_{ij}}=-\nabla_i T_\mathrm{div}^{ij}=0~.
\ee
If there are matter fields, apart from the additional Ward
identities related to other local symmetries, even the diffeomorphism identities 
will be modified. In fact, in the presence of nonzero vacuum expectation values 
the CFT stress-energy tensor is not conserved, but obeys rather,  for one scalar
source, say, 
\be
\nabla^j \<T_{ij}>= -\<{\cal O}> \nabla_i J~.
\ee
This should not be considered, however, as a diffeomorphism anomaly, but rather 
a feature of the spontaneous symmetry breaking
induced by the nonzero vacuum expectation values of ${\cal O}$. 
This is in accord with the non-conservation of the Brown--York 
\cite{quasilocal} quasi-local stress-energy tensor, as was previously noticed 
in \cite{Skenderis1}. 
The signature of a gravitational anomaly is, instead, 
stress-energy nonconservation with no vev's turned on, analogously to what happens
if there is a holographic chiral anomaly \cite{Witten1,holoWard}.
For obtaining an explicit anomaly in pure gravity it would therefore 
be necessary to consider the appropriate Chern--Simons
terms. Such a term would be, for instance, 
in the $(2,0)$ theory in 6 dimensions the reduction on $S^4$ of the M-theory CS 
term 
\be
S_\mathrm{CS}=\int \dd^7 x \Tr (\omega \wedge R \wedge R \wedge R)~.
\ee
We shall see in Section \ref{form} that in the presence of fermion fields 
this pattern will be somewhat modified.

\section{The Holographic RG Equations}

In this section we review how the holographic RG equations arise. Here, 
we do not restrict 
to pure gravity theories, but assume that the brackets in (\ref{CS1})  
 have been modified to accommodate variations with respect to  
matter fields. Equation (\ref{CS1})  reads
\be 
(S_\mathrm{div},S_\mathrm{div})+2(S_\mathrm{div},W)+
(W,W)={\cal L}~.\label{exactHRG}
\ee  
The second term on the LHS is a linear operator acting on $W$.
$(W,W)$ is a quadratic 
correction to the linearized RG, and the left-over piece gives additional
non-homogeneous terms, whose leading contribution was shown in the
previous section to give the Weyl anomaly. 
In fact, because of cancellations
achieved in the descent equations (\ref{fourth}), 
(\ref{exactHRG}) can be rewritten 
\beq
2(S_\mathrm{div},W)+(W,W)={\cal O}(1)~.\label{HRG}
\eeq
In order to get a Callan--Symanzik type equation, one can take the following 
steps \cite{dBVV}: 
Calculate $n$ variations of (\ref{HRG}) w.r.t.~the sources and drop the
contact terms containing more then one delta function, which would not
contribute for operators evaluated at different points. 
Then letting the sources go to constant (in $x_i$) values, the metric 
assuming the form,
\beq 
g_{ij}=\mu (t)\hat{\eta}_{ij}~,\label{scaletransf} 
\eeq
and eventually integrating once over $\dd x^d$ one gets expressions of the form 
\be
\left( {\cal U}\mu\frac{\de}{\de \mu}+ \beta_I \frac{\de }{\de J_I}\right)
\<{\cal O}_{I_1}{\cal O}_{I_2}>   
+\sum_{n=1}^2 \de_{I_i} \beta^{J_i} 
\<{\cal O}_I {\cal O}_{J_i}>={\cal O}(t^d\log^2 t)~,\label{intermediateHRG}
\ee  
where\footnote{We drop a numerical coefficient.}
\be
\beta_I & = & \left(\frac{1}{\sqrt{\hat{g}}}
G_{IJ}\frac{\delta S_\mathrm{div}}{\delta J_J}\right)_{J=J(t),g=
\hat{\eta}}\\
{\cal U} & \simeq &\left(\frac{1}{\sqrt{\hat{g}}} 
g_{ij}\frac{\delta S_\mathrm{div}}{\delta g_{ij}}\right)_{J=J(t),
g=\hat{\eta}}~.
\ee
Notice that these equations are taken at a finite cutoff $t$, 
{\it i.e.} for {\it bare} quantities.
Would one allow the couplings to be space-time dependent the 
vanishing of the beta functions would correspond to the sources obeying 
$d$-dimensional Einstein equations that would follow from  $S_\mathrm{div}$.
This could give rise in principle to boundary backgrounds that {\it are not}
Minkowski, or its conformal completion.  

Considering finally the Hamiltonian flow equations (\ref{flow}) 
it turns out that in the appropriate gauge the transverse coordinate plays 
the role of a parameter for the scale transformation induced by the metric.  
In general it is not always true that a gravity solution has the interpretation 
of a RG flow: It depends in fact on
the leading behaviour of the bulk fields near the boundary of AdS
\cite{Balasubraplusminus,KWplusminus}. In the
Hamilton--Jacobi context this means that not every solution $S_\mathrm{div}$ of
the constraints
gives rise to physical flows \cite{Martelli}. However, for those that are physical, 
the flow equations read, after fixing the gauge $N_i=0$
\be
\dot{\mu} & \simeq &  N~{\cal U}~\mu \label{scale}\\
\dot{J_I} & \simeq &  N~\beta_I~.\label{scalarflow} 
\ee 
The scale transformation depends parametrically on the
cutoff $t$. Let us now use (\ref{scale}) to express (\ref{intermediateHRG}) 
in terms of the variation of $t$, 
and choose the gauge $N=+\frac{1}{t}$. This yields 
\be
(t\frac{\de}{\de t}+\beta_I\frac{\de}{\de J_I})\<{\cal O}_{I_1}
\cdots {\cal O}_{I_n}> \nonumber & & \\ 
+\sum_{i=1}^n \de_{I_i} \beta^{J_i} \<{\cal O}_{I_1}\cdots
{\cal O}_{J_i} \cdots {\cal O}_{I_n}> &=& {\cal O}(t^d\log^2 t)~,
\label{bareHRG}
\ee
and (\ref{scalarflow}) can be written as 
\be
\beta_I & = & t\frac{\dd ~}{\dd t} J_I~,
\ee
which is consistent with the beta functions defined before and with 
interpreting (\ref{bareHRG}) as a bare RG equation. Furthermore the 
right hand side represents higher order terms that, in this spirit,
are logarithmic corrections to scaling.
This choice of gauge differs from the Fefferman--Graham \cite{FG}
by a sign: The latter corresponds in fact to $N=-\frac{1}{t}$. 

As is the case in QFT, the definition of the beta
functions is not unique. They are fixed unambiguously near the 
fixed points, where their behaviour is
universal, but their extrapolation at intermediate RG steps is not 
\cite{Cfunction}. The definition given in \cite{dBVV} differs
from that given above by the ratio of the rate of scale change (\ref{scale}).
In fact, in that reference  $N=1$ and  all quantities depend on $\mu$. 
The fixed points are not affected, because the zeros of the beta
function cannot be modified by dividing by  ${\cal U}$. 
Also, it is well know that the bare RG equations and the renormalized ones 
have the same physical content.

\section{Fermions and Form Fields in Hamiltonian Theory}

Let us now turn to a generic $(d+1)$-dimensional theory with spinors, 
form fields and local diffeomorphism invariance. 
We will consider the most general two-derivative action (neglecting
Chern--Simons terms)
with quadratic fermion couplings consistent with gauge symmetry, namely 
$S = S_I + S_{II} + S_{III}$, where
\be
S_I &=& \frac{1}{\kappa^2} \int \dd^{d+1} x \sqrt{g} \left(  \tilde{R} 
- 2\eta \tilde\nabla\cdot\tilde\nabla_n n + 2\eta 
\tilde\nabla \cdot (n \tr K)  - \kappa^2 \Lambda \right)  \label{sI} \\
S_{II} &=&  \int \dd^{d+1} x \sqrt{g} \left( \frac{1}{2 \lambda^2 } F_A F^A + F_A J^A \right) \\
S_{III} &=&  \half \int \dd^{d+1}x~ \sqrt{ g} \left( {\bar\psi}
M \Dslash \psi - (\Dslash{\bar\psi})M\psi + 
2 \bar{\psi} Z_A \Gamma^A \psi \right) ~.
\ee
The fields appearing in these expressions are the following: 
The field strength $F = \dd A$ is an Abelian $p$-form and 
couples to the fermions through $J^A = \bar\psi \zeta \Gamma^A \psi$. 
The capital Latin letters refer to a multi-index of pertinent rank;  
summations include division by the factorial of the rank.
There can be arbitrarily many fermion flavours, 
but we always suppress the index that would distinguish them. 
This action encaptures, and generalizes, many 
interesting features of the effective superstring actions. For instance, $F$ 
could be thought of as a Ramond--Ramond field.

The nondynamical couplings $\zeta$, $Z_A$ and $M$ 
mix fermion flavours, and are not assumed space-time constants, 
unless explicitly indicated. They satisfy suitable hermiticity 
conditions so that the action is always real.
They can be thought of as the Yukawa couplings to 
higgsed scalar fields. As $Z_A$ is not a dynamical field, we have assumed $Z_{0A} =0$.  
The underlying space-time has a boundary to which we can associate 
a normal vector field $n$ and the extrinsic curvature $K$. This vector field has constant 
norm  $n \cdot n = \eta = \pm 1$, and is hence either temporal or spatial. 

In order to go over to the Hamiltonian formalism, 
we need to choose a particular ``time coordinate'', or ``evolution parameter'' 
whose tangent field we call $e_0$. The bulk vielbeine 
$e_\mu^{~~\ua}$ decompose into boundary vielbeine $L_i^{~~\ua}$, the vector
$n$, and the 
nondynamical degrees of freedom $N$ and $N^i$. The connection on the boundary  
is obtained (see Appendix
\ref{notation}) from that in the bulk by shifting the 
Christoffel symbols in such a way that 
the $n$ becomes a covariantly constant vector field on the boundary, and that the 
boundary basis is covariantly constant in the normal direction
\be
\nabla_i ~n & = & 0\\
n\cdot \nabla_i ~e_j & = &0~.
\ee

Given the time direction $e_0$ we can now reduce all tensor fields in terms
of fields defined on the boundary. Isolating the kinetic terms in the action,
{\it i.e.} terms involving derivatives along $e_0\,$, the remainder is by 
definition the total Hamiltonian 
\be
S &=&  \int \dd t~ \left( p~ \dot{q} - H \right)  
\ee
where
\be 
p~ \dot{q} &=& 
\int \dd^{d} x \left( p^{i\ua}~ \partial_t L_{i\ua}  + E^{\hat{A}}~ 
\partial_t A_{\hat{A}} + \bar\chi~ \partial_t \psi -
 \partial_t \bar\psi~ \chi  \right) \\
H &=&  \int \dd^{d} x \left( N {\cal H}_\perp + N^i {\cal H}_i + 
A_0^{\hat{A}} {\cal G}_{\hat{A}} + \varepsilon_{\uab} {\cal J}^{\uab} \right)
\ee
We have added here, by hand, the constraint that guarantees the freedom to 
choose the flat basis for vielbeine freely, {\it i.e.} the generator of local
Lorentz transformations in 
the bulk, ${\cal J}^{\uab}$. For notation, that is standard, see 
Appendix \ref{notation}.

The physical phase space can now be easily read off from the 
above reformulation. It consists of the canonical pairs $(p^i_\ua, L_j^\ub)$, 
$(E^{\hat{A}}, A_{\hat{B}})$, $(\bar\chi^a, \psi_b)$, and 
$(\bar\psi^a, \chi_b)$. 
The fields $A_0^{\hat{A}}\,$, $N$ and $N^i$ are Lagrange multipliers that
correspond to the first class constraints 
${\cal G}_{\hat{A}}\,$, ${\cal H}_\perp$ and ${\cal H}_i$, respectively.
Due to fermions, that have a first order action principle, there are also 
{second class constraints}
\be
\chi &=& \half  \eta\sqrt{\hat{g}}~ \Gamma^n M \psi~.
\ee
In order to solve these constraints, we have to split the fermion phase space 
in some Lorentz invariant way into two parts, one of which we treat as
the configuration space, and hence boundary fields, and the other of which  are then 
the momenta, to be solved as functionals of the boundary fields in HJ theory. 
This is done in Appendix \ref{fermi} by imposing a chirality condition;
here we shall only point out some salient features. 

Most of the calculations must be done case by case, 
but the emerging structures are similar. Having used the 
condition of choice to put the kinetic term in such a 
form (as in formulae (\ref{kinfermi}) and  (\ref{kin2})) 
that the coordinates and the momenta can be read off, 
there are always some left-over pieces:  
\begin{itemize}
\item[1)] There turns out to be a total time-derivative term $\de_t G$.
This term should be simply subtracted from the action, 
as argued in \cite{spinors3,spinors4}. 
Generally speaking the reason for this procedure is that the generating function arises also 
on the bulk side, eventually, from a path integral, which is going to be defined more 
fundamentally in Hamiltonian language. 
It has also been pointed out  \cite{spinors1} that, as first order actions vanish 
on classical solutions, we would otherwise 
get a trivial result as far as the fermions are concerned. 
\item[2)]
There will also arise terms that involve a time 
derivative of the metric. This will naturally 
change the gravitational momentum, 
but in a way that is easily kept track of. After
some algebra it turns out to be sufficient 
to simply shift  by
\be
\pi^{ij} &\longrightarrow& \pi^{ij} - \half \hat{g}^{ij} G~. 
\ee
\end{itemize}

\section{The Holographic Callan--Symanzik Equation}
\label{bracket}

Having split the fermion phase space we are now in the position to
write down the Hamilton--Jacobi equations for the full system. 
We will perform this analysis for Weyl fermions and  assume that  $M$ 
is a space-time constant matrix. 

It turns out that, provided there are no marginal operators $Z$ present 
in the bulk and that the rank of the tensor field $p$ is {\em odd},
the Hamilton--Jacobi equation originating from ${\cal H}_\perp$ does indeed 
take the  generally expected form 
\be
(F,F) = {\cal L}~\label{bracketeq}. 
\ee
 Where now 
\be 
(F,F) &=& (F,F)_g + (F,F)_A + (F,F)_\varphi~,
\ee
and the RHS of  (\ref{bracketeq}) is 
\be
{\cal L} &=& \sqrt{\hat{g}} \left( \frac{1}{\kappa^2}  R - \Lambda 
 + \frac{1}{2 \lambda^2 } F_{\hat{A}} F^{\hat{A}} + 
F_{\hat{A}}  \bar\varphi \zeta \Gamma^{\hat{A}} \varphi
 \nonumber \right. \\ 
& & \left. \qquad \qquad + \half  {\bar\varphi}
M \hat{\Dslash} \varphi - \half (\hat{\Dslash}{\bar\varphi})M\varphi + 
 \bar{\varphi} Z_{\hat{A}} \Gamma^{\hat{A}} \varphi \right)~. \label{L}
\ee
It is useful to define the following 
operators\footnote{Derivatives act from the left, but not on fields included in the same operator.}
\be
{\cal D} &=& \half \bar\varphi \frac{\delta}{\delta \bar\varphi} 
-  \half\frac{\delta}{\delta \varphi} \varphi \label{bigd} \\
{\cal D}^{ij} &=&  \frac{\delta}{\delta g_{ij}} - \half \hat{g}^{ij} {\cal
D} \\
{\cal D}^{\hat{A}} &=& \frac{\delta}{\delta A_{\hat{A}}} +
\bar\varphi \zeta M^{-1} \Gamma^{\hat{A}} \frac{\delta}{\delta \bar\varphi} + 
\frac{\delta}{\delta \varphi} M^{-1}\zeta \Gamma^{\hat{A}} \varphi
 \ee
Had we also considered $p$ even, 
the last equation would have been different: Then the operator ${\cal D}^{\hat{A}}$ 
would have contained terms with either no or two derivatives w.r.t.~the fermion fields. 
Now the brackets can be written easily\footnote{Here we have set $Z$ to
zero. See below.} 
\be
(F,H)_g &=& \frac{-\eta \kappa^2}{\sqrt{\hat{g}}} (g_{il}g_{jk} - \frac{1}{d-1} g_{ij} g_{kl})~   
({\cal D}^{ij} F)~ ({\cal D}^{kl} H) \\
(F,H)_A &=& \frac{\eta \lambda^2}{2 \sqrt{\hat{g}}}~  
({\cal D}^{\hat{A}} F)~ ({\cal D}_{\hat{A}} H) \\
(F,H)_\varphi &=& \frac{-\eta}{2 \sqrt{\hat{g}}}~  \left( 
\frac{\delta F}{\delta \varphi} M^{-1} \hat{\Dslash} \frac{\delta H}{\delta 
\bar\varphi} 
- (\hat{\Dslash} \frac{\delta F}{\delta \varphi}) M^{-1} \frac{\delta H}{\delta \bar\varphi}
 \right)~.
\ee
The reason for the fact that also the fermionic momenta give rise to a bracket 
is easily seen in the case for Weyl fermions: The chiralities of the coordinates 
and the momenta are such that if we insert any operator even in Clifford matrices 
between them,  $\bar\pi {\cal O}_{\mathrm{even}} \varphi$, 
the result is nontrivial. Similarly, the nontrivial results
for odd operators arise from insertions between either two coordinates, $\bar\varphi {\cal O}_{\mathrm{odd}} \varphi$,   or two momenta, $\bar\pi {\cal O}_{\mathrm{odd}} \pi$.

This structure changes slightly if the bulk mass terms, 
or couplings to external form fields, $Z$, are included. 
If $Z$ is even, such as a mass term, there will be an additional  term
\be
\delta_Z F + (F,F) = {\cal L}
\ee
where
\be
\delta_Z &=&  \frac{1}{\sqrt{\eta}} 
\left( 
\bar\varphi Z M^{-1} \frac{\delta}{\delta \bar\varphi} -
\frac{\delta}{\delta \varphi} M^{-1} Z  \varphi 
\right)~.
\ee
This addition preserves the flow equation 
form of the final Callan--Symanzik equations, however. 
Also, if  $Z$ is odd or $M$ is not constant, there will be an additional 
quadratic piece in the fermion momenta, and the basic form of the brackets 
will be, eventually, unchanged.

If we are forced 
to give the initial data in terms of Weyl fermions, 
as we are assuming in this article, 
the requirement that the bulk action produce 
a QFT generating functional that obeys the Callan--Symanzik 
equation implies that bulk theories 
where fermions are coupled to dynamical {\em even} rank form fields 
do not posses a simple holographic dual.

Including higher order interactions could be a problem, because they would 
introduce higher powers of momenta, which would spoil the basic form of the 
brackets. 
However, we have seen above  an encouraging rearrangement of terms,  where
the structure of the theory solves a similar problem. For instance, the form field kinetic 
term absorbs some four fermion couplings in the expression 
$({\cal D}^{\hat{A}} F)^2$. We can indeed view the fermionic additions 
in ${\cal D}^{\hat{A}}$ as a covariantization of 
the flat derivative with respect to the form field $\hat{A}$. There is, therefore, reason to expect 
that at least in theories that are known to have a holographic dual but which contain 
four fermion interactions the additional symmetries, such as local supersymmetry or the
$SL(2,\mathbb{Z})$ invariance in type IIB SUGRA, 
might arrange the fermion structure in such a way that the eventual higher fermion derivatives 
would still be manageable.

\section{First Class Constraints as Ward Identities}
\label{form}

In addition to the constraint ${\cal H}_\perp$ treated above, 
we have to solve also the rest of the first class constraints
${\cal G}_{\hat{A}}$, ${\cal H}_i$ and ${\cal J}^{\uab}$. 
Generally speaking, they will impose gauge symmetry, 
diffeomorphism invariance and local Lorentz symmetry on the boundary.
Why this is not straight forwardly so is because, in the bulk, these constraints 
generate symmetry transformations through Dirac brackets. 
Due to the ansatz (\ref{pqF}) in Hamilton--Jacobi formalism, they will 
act differently on $F$, thus resulting in conditions that do not impose
necessarily the full off-shell symmetries of the bulk theory. 
As far as the Lorentz and the gauge symmetries are concerned the geometrically expected
actions are obtained. In case the of diffeomorphism invariance some additional 
constraints arise.

For instance, the fact that $A_{\hat{A}}$ enters the generating functional 
$F$ only 
through its field strength is sufficient to guarantee ${\cal
G}_{\hat{A}}=0$. 
This simply reflects the fact that the boundary theory must
have the same gauge symmetry as the bulk theory. A similar situation 
prevails as far as ${\cal J}^\uab$ is concerned:  
$n_\ua {\cal J}^\uab =0$ just because $n_\ua L_i^{~~\ua} =0$, and the rest of the components 
generate the expected action on vielbeine and spinors, through
\be
\Lambda^{jk} &=& L^{[j}_{~\ua} \frac{\delta}{\delta L_{k]\ua}} - 
\frac{1}{4} \left(  \frac{\delta}{\delta \varphi} \Gamma^{jk}\varphi +  
\bar\varphi \Gamma^{jk} \frac{\delta}{\delta \bar\varphi} \right)~.
\ee
This constraint guarantees, therefore, local Lorentz invariance on the boundary. Choosing gauge
and Lorentz invariant $S_{\mathrm{div}}$ will be enough to avoid anomalies in the boundary theory.

The situation is somewhat more involved when the constraints that guarantee
diffeomorphism invariance are considered.
This would mean that the effective action 
be invariant under translations generated by a 
vector field $\chi$. Or, in other words, that shifting the fields $\hat{A}$
and $\varphi$ infinitesimally by their Lie-derivatives 
\be
 {\cal L}_{\chi} \hat{A} &=& {\iota}_{\chi} \hat{F} + \dd {\iota}_{\chi}
 \hat{A} \label{LA}\\
 {\cal L}_{\chi} \varphi &=& (\hat{D}_{\chi} + \frac{1}{4} \nabla_{[i} 
\chi_{j]} \Gamma^{ij} ) \varphi \label{Lphi}
\ee
we get a contribution that combines 
together with a contribution from the integration measure 
to a total derivative of the Lagrangian. 
Note that a general spinorial Lie-derivative does not obey the Leibniz 
rule\footnote{For a general introduction to spinors and geometry see for
instance \cite{book}.}. 
It is therefore useful to restrict to Killing fields $\nabla_{(i} \chi_{j)} =0$,
for which the formula (\ref{Lphi}) applies. 

Solving the constraint ${\cal H}_i =0$ we get, again, assuming $p$ odd and 
the Weyl decomposition, that the variation
\be
{\delta}_{\chi} &=&  \nabla_k\chi^j~ L_{j\ua} \frac{\delta}{\delta L_{k\ua}} +
\chi^i \left( \partial_{[i} A_{\hat{A}]} {\cal D}^{\hat{A}} - 
  \frac{\delta}{\delta \varphi} \hat{D}_i \varphi +
 \hat{D}_i \bar\varphi  \frac{\delta}{\delta \bar\varphi} \right) \label{DI}
\ee
should annihilate the effective action. This differs from a 
Lie-derivative in two respects: First, the transformation of the form field 
is accompanied by a gauge transformation. Second, its action on fermions is 
modified by  
\be
\Delta_\chi \varphi &=& - \iota_\chi F_{\hat{A}}~ \Gamma^{\hat{A}} M^{-1} \zeta  \varphi  \\
\Delta_\chi \bar\varphi&=& \iota_\chi F_{\hat{A}}~ \bar\varphi \Gamma^{\hat{A}} \zeta M^{-1}
\ee
where $\Delta_\chi =\delta_\chi - {\cal L}_\chi$. 
If we want to restrict to 
theories where the boundary diffeomorphism invariance 
still prevails, we have to put this difference
to zero. 
A solution of $\Delta_\chi S_{\mathrm{div}} =0$ is ensured imposing the following conditions:
\begin{itemize}
\item[1)] The Clifford action of any differential form $K$ 
appearing in the fermion couplings  $\bar\varphi K_{\hat{A}} \Gamma^{\hat{A}} \varphi$ 
commute with  that of ${\iota_{\chi}} \hat{F}$, {\em i.e.}
\be
[ K_{\hat{A}} \Gamma^{\hat{A}}, \chi^i F_{i\hat{B}} \Gamma^{\hat{B}}] = 0~. \label{cc1}
\ee
For instance, for $K_{\hat{A}}=F_{\hat{A}}$, $F_{\hat{A}}$ being
of odd rank,  this is clearly true. This condition means then, geometrically, that $K_{\hat{A}}$ and 
$F_{\hat{A}}$ should be aligned in a certain way.
In addition to this the couplings should satisfy, in the notation of formula (\ref{ansatz}), 
\be
\hat\zeta M^{-1} \zeta = \zeta M^{-1} \hat\zeta~.
\ee
\item[2)] If there is a kinetic term on the boundary, such as $\bar\varphi \hat{\Dslash} \varphi -  
(\hat{\Dslash} \bar\varphi)  \varphi$ the following restrictions be true
\be
{\cal L}_\chi \hat{F} &=& 0 \\
 \iota_\chi F^{i\hat{A}} \hat{D}_i \varphi &=& 0 
\ee
These are strong requirements, as they concern the 
boundary fields and not only couplings, and therefore really restrict,
from the bulk point of view, 
the set of acceptable initial conditions. The simplest way to solve them is naturally 
to exclude the kinetic terms from the action, cf.~end of Section \ref{locexp}. 
However, this might be too drastic a solution, 
as, in the above, we are assuming that there exist a
Killing field $\chi$; after all, we are considering an interacting theory
with physical sources.
More interestingly, these conditions can be solved by assuming
$\iota_\chi \hat{F}=0$: This means that the Killing isometry only changes the field 
$\hat{A}$ by generating a gauge transformation. With this understanding it is 
sufficient to set    
\be
{\cal L}_\chi \hat{A} \approx 0~. \label{symm} 
\ee
This would mean that there are no restrictions on the fermion fields, 
whereas the form field potential is frozen to configurations covariant 
under flows generated by $\chi$.
\end{itemize}
Then, considering diffeomorphism invariant terms in $S_{\mathrm{div}}$, one obtains
the Ward identity
\be
{\cal L}_\chi W  + \Delta_\chi \bar\varphi \langle {\cal O}_{\bar\varphi} \rangle 
- \langle {\cal O}_\varphi \rangle \Delta_\chi \varphi 
  &=& - \Delta_\chi S_{\mathrm{div}}~.
\ee
The RHS is the failure of the counter terms to satisfy the constraint 
${\cal H}_i = 0$, 
while the vacuum expectation values would signal the spontaneous symmetry breaking of 
Lorentz symmetry in the dual QFT. This is analogous to the analysis in 
Sec.~\ref{RA} where scalar fields coupled to gravity  were considered.

\section{Local Expansion in the Bracket Equation}
\label{locexp}

Let us consider a particular solution of the classical 
equations of motion that behaves at the boundary $t\longrightarrow 0$ 
as
\be
g_{ij}(x,t) &=& t^{-2} g_{ij}^R(x) + {\cal O}(t^{-1}) \\
A_{\hat{A}}(x,t) &=& t^\nu A_{\hat{A}}^R(x) + {\cal O}(t^{\nu+1}) \\
\psi(x,t) &=& t^\sigma \psi^R(x) + {\cal O}(t^{\sigma+1})
\ee
in the coordinate system of Eq.~(\ref{rsquare}).
The quantities with superscript {\footnotesize $R$} refer to expressions 
that have a finite and nonzero limit at $t \longrightarrow 0$.
We will actually not need to show whether the full coupled system has 
solutions with this particular asymptotic behaviour. 
This would also be quite difficult --- 
it was found, for instance, in \cite{spinors3,spinors4} that free
fermions scale as
$\sigma= d/2 -m$, where $m$ is the bulk mass. In our case the coupling 
of the form field $F_A$
to the fermions gives rise to an effective mass term. 
We cannot, however, fix 
the field $F_A$ in any useful way in order to analyse conclusively 
the scaling behaviour in this coupled system.

Instead, we impose the above scaling behaviour for some set of critical exponents
$\nu$ and $\sigma$ and then derive from this -- and the assumption that the
holographic dual exist at all 
-- consistency conditions for both the bulk and the boundary theories. 
Let us consider, in particular, the case $\nu = -n + 1$ and $\sigma=1/2$. 
This assignment of critical exponents has the virtue that the expansion of
$S_\mathrm{div}$ will look like an expansion in terms of the naive mass
dimension relevant to supersymmetric models. This choice will turn out to be
a convenient book-keeping device, but the results we eventually get are
valid more generally. In order to solve the full bracket equation,
 we will have to arrange the coefficients of terms
that not only have the same scaling behaviour, but also the same structure, to cancel.
So, in principle, one can check {\it a posteriori} for which range of scaling 
exponents the terms we neglected are still subleading and our results continue to be 
then valid. 
We shall further assume that all couplings 
such as $M$ and $Z$ are marginal operators, and as such $t$-independent. This
assumption is not very restrictive,  not even in the presence of scalar fields, 
that would render the couplings dynamical.
We can now write a local ansatz for $S_\mathrm{div}$ that is of the same functional form as (\ref{L})
\be
S_\mathrm{div} &=&  \int \dd^{d} x \sqrt{\hat{g}} \left( k_1 R - \hat{\Lambda}  
 + \half k_2 F_{\hat{A}} F^{\hat{A}} + F_{\hat{A}}  \bar\varphi \hat{\zeta} \Gamma^{\hat{A}} \varphi
 \nonumber \right. \\ 
& & \left. \qquad \qquad + \half  {\bar\varphi}
\hat{M} \hat{\Dslash} \varphi - \half (\hat{\Dslash}{\bar\varphi}) \hat{M} \varphi + 
 \bar{\varphi} L_{\hat{A}} \Gamma^{\hat{A}} \varphi \right)~. \label{ansatz}
\ee
As we shall later restrict to a scalar coupling $L$, it turns out that 
no four fermion terms are needed.
The couplings $k_1, k_2, \hat{\zeta},\hat{M}$ and $L_{\hat{A}} \Gamma^{\hat{A}}$ need not 
be constant on the boundary, but they are assumed marginal, {\em i.e.}~time-independent.

At leading order $S_\mathrm{div}$ will diverge as $t^{-d}$ when 
$t\longrightarrow 0$. We shall consider the three lowest order 
contributions to the equation ${\cal H}_\perp =0$, or, 
$(S_\mathrm{div},S_\mathrm{div}) = {\cal L}$.
\begin{itemize}
\item[1)]
At the leading order we get 
\be
\Lambda + \frac{\eta \kappa^2}{4} \frac{d}{d-1} \hat{\Lambda}^2 = 0~.
\label{LL}
\ee
The boundary cosmological constant is therefore essentially 
the square root of the bulk one and sets the scale.
We notice that depending on the sign of the bulk gravitational constant
we can choose to look at either time-like or space-like boundary surfaces.
However, only the space-like surfaces admit an asymptotically anti de Sitter solution.
\item[2)]
The  ${\cal O}(t^{-d+1})$ equations depend on the details of $Z$. The equations simplify
assuming that $Z$ is odd in Clifford matrices and 
that $L$ is a scalar, {\em i.e.}~a flavour matrix, as we then get
\be
0 &=& Z  +  \eta L M^{-1} Z M^{-1} L \label{v0}~.
\ee
If $Z$ is even, we get
\be
Z M^{-1} L + L M^{-1} Z &=& 0~.
\ee
In particular, we could consistently set $L=0$. This would mean that, for instance, 
a bulk mass term does not automatically lead to a mass term on 
the boundary. 
\item[3)]
At the next order  ${\cal O}(t^{-d+2})$ we get 
from the Einstein--Hilbert and  the form field kinetic terms 
conditions for the couplings $k_1$ and $k_2$
\be
\frac{\eta}{d-1}  \kappa^2\hat{\Lambda} (\frac{d}{2} - 1) k_1
-\frac{1}{\kappa^2} = 0 \label{v1} \\
\frac{\eta}{d-1}  \kappa^2\hat{\Lambda} (\frac{d}{2} - p) k_2 
-\frac{1}{\lambda^2} = 0 \label{v1.2} ~.
\ee
For $d>2$ equation (\ref{v1}) allows one to compute 
the coefficient of the $R$ term, as
expected. In $d=2$ this equation does not arise as $\sqrt{\hat{g}}R$ is
marginal, and it is not to be included in $S_\mathrm{div}$. Instead,
the impossibility of canceling it in the descent equations translates
to the Weyl anomaly in $d=2$ as in Eq.~(\ref{dtw}). 

Equation (\ref{v1.2}) gives the value of $k_2$ for $d\neq 2p$. Note, however, 
that for middle-dimensional form fields $d=2p$ this equation will not
have solutions. In the case of free form fields, as they are always 
marginal \cite{marika}, it is just the anolgue of (\ref{v1}) for pure gravity:
In this case $F^{\hat{A}} F_{\hat{A}}$ is a marginal operator and it contributes to the matter part of the 
Weyl anomaly. In the interacting case the 
treatment of this term depends on the scaling dimension $\nu$.
If $\nu <0$ then the contribution should really be included in the divergent part, and 
a middle-dimensional form field would not allow consistent solutions for the bracket equation.

\item[4)]
Assuming that $M$ and $L$ are constants on the boundary, 
the fermion terms yield the constraints
\be
-\frac{p}{2} \frac{\eta}{d-1}  \kappa^2\hat{\Lambda} \hat{M}  &=& M + \eta L M^{-1} L \\
-\frac{\eta}{2d-2}  \kappa^2\hat{\Lambda} \hat{\zeta} &=& \zeta +  
\eta L M^{-1} \zeta M^{-1} L \label{v4}~.
\ee
\end{itemize}
In all of the equations (\ref{LL}) and (\ref{v1}) -- (\ref{v4}) we see that the relationship 
between the bulk and the boundary couplings is essentially 
a scale factor $\kappa^2\hat{\Lambda}$. It is interesting to note that the role of 
$L$ is to mix bulk fermion flavours into new combinations on the boundary. Assuming 
$L$ proportional to $M$ would lead to scaling the fermion field. Furthermore, setting 
$L=\sqrt{-\eta} M$ 
the dynamical part to the fermion action drops out completely $\hat{M} = 
\hat{\zeta}=0$, and the only contribution comes from the $L$-term itself 
\be
\sqrt{-\eta} \int \dd^d x \sqrt{\hat{g}} \bar\varphi M \varphi~.
\label{suitable}
\ee
As discussed in the previous section this solution 
would not produce a diffeomorphism anomaly,
even without restricting to symmetric form fields, cf.~Eq.~(\ref{symm}).

\section{Conclusions}

We have investigated the relationship between diffeomorphism 
invariant theories and their holographic duals, showing in particular that, 
in a theory that contains fermions, nontrivial consistency conditions arise. 
These conditions restrict, for instance, the couplings of even rank form field
field strengths to fermions. 

After explaining how the Hamiltonian reduction is performed, with particular
attention to fermion fields, in Sections \ref{bracket} and \ref{form} we have 
discussed how the holographic Callan--Symanzik equation and other Ward 
identities following from first class constraints get modified in the presence
of fermions and forms.
Although the gauge and the Lorentz constraints did not lead to surprises, the  
Poincar\'e constraints resulted in an anomalous contribution 
in the diffeomorphism Ward identity on the boundary. 
However, one can get rid of these terms, either imposing conditions on the 
sources, as for instance the boundary gauge potential to be
constant in the sense of Eq. (\ref{symm}), or by the choice of a suitable
counter term (\ref{suitable}). 

We have also derived relationships between the bulk and boundary couplings, 
finding that the role of the cosmological constant on the boundary 
is to set the scale of all boundary couplings w.r.t~bulk couplings, as
expected. Moreover, we were able to find the first terms of the expansion
of the counter term action $S_\mathrm{div}$, with couplings fixed
in terms of those appearing in the bulk action. This analysis shows also
that one can avoid including fermion kinetic terms
in $S_\mathrm{div}$ that would have a nonzero anomalous diffeomorphism 
variation.

There still remain  open problems. 
One should clarify, for instance, what kind of restrictions higher fermion couplings 
impose on the duals, or what role  other 
bulk symmetries, such as supersymmetry, might play in the bracket structure. In particular,
it would be interesting to include dynamical scalar fields and Rarita--Schwinger fermions 
in these considerations.
One would also like to know how robust our results actually are when the scaling behaviour 
of the bulk fields at the boundary are varied. Finally, the formal structures 
arising here are quite intriguing: It would be interesting to find out whether 
the $(\cdot,\cdot)$-bracket or the ${\cal D}$-operators have
a geometrical meaning in solving Ward identities.

\vspace{2cm}

\noindent
{\bf Acknowledgements}: We thank C.~Acatrinei, M.~Henningson, R.~Iengo, and A.~Mukherjee
for discussions. We are grateful
to G.~Ferretti for useful comments and encouragement. D.M.~would like to thank
the department of Physics of Chalmers University (G\"oteborg) for hospitality
during the latest stages of this work. D.M.~has received financial support
from Chalmers University of G\"oteborg.
This work was supported in part by the EC TMR program CT960045.

\appendix

\section{Notation and useful formulae}
\label{notation}

Greek indices $\mu,\nu,\ldots$ refer to the coordinate directions in the 
bulk, underlined Greek indices $\ua,\ub,\ldots$ are flat indices in the bulk,
lower case Latin indices $i,j,\ldots$ refer to coordinate directions in the boundary,
upper case Latin indices $A,B,\ldots$ refer to normalized multi-indices in the bulk, and
hatted upper case Latin indices $\hat{A},\hat{B},\ldots$ 
refer to normalized multi-indices in the boundary. If there is danger of confusion, 
 symbols with a tilde, such as $\tilde\nabla$, are used to refer to bulk quantities 
and symbols with a hat, such as $\hat\Dslash$, to boundary quantities.

We can express the connection between an arbitrary flat 
coordinate basis $\{e_\ua\}$ and a basis $\{e_0, e_i\}$
that involves the direction of the evolution coordinate (bulk direction) $e_0$ 
in terms of the vielbeine
\be
e_i^{~~\ua} &=& L_i^{~~\ua} \\
e_0^{~~\ua} &=& N n^\ua +N^i L_i^{~~\ua}~.
\ee
Due to the algebraic constraint $L_i^{~~\ua} n_\ua =0$ we have 
$e_i \cdot n =0$. For other properties of this frame see for instance 
\cite{Nelson&Teitelboim}. 
The boundary metric and vielbeine are related through
\be
g_{ij} & = &  L_i^{~~\ua} L_j^{~~\ub} \eta_{\uab}\\ 
\eta_{\uab} &=&  L_{i\ua} L^i_{~\ub} + \eta n_\ua n_\ub~, 
\ee
and the boundary gamma matrices are defined by 
\be
\Gamma^n & =  & n_{\ua} \Gamma^\ua \\  
\Gamma^i & = & L^i_{~\ua} \Gamma^\ua~.
\ee

Given the Levi--Civita  connection $\tilde\nabla$ in the bulk
we can construct a metric connection on the boundary by setting \cite{Diaz} 
\be
\nabla_X Y &=& \tilde\nabla_X Y + \eta n (Y \cdot \tilde\nabla_X n ) - 
\eta \tilde\nabla_X n (Y \cdot n )
\ee
for arbitrary vector fields $X,Y$. This connection enjoys the properties
\be
\nabla_i n &=& 0 \\
n \cdot \nabla_i e_j &=& 0~.
\ee
The spin connection in the bulk can be expressed in terms of that on the boundary using  
\be
\tilde\omega_{i\uab} &=& \Gamma_{jik}~ L^{j}_\ua~ L^{k}_\ub -  L^{k}_\ub~ \partial_i L_{k\ua} - \eta~  
n_\ub~ \partial_i n_{\ua} \nonumber  \\
& & + \eta~ K_{ij}~ (n_\ua~ L^{j}_\ub -  n_\ub~ L^{j}_\ua ) \\
\tilde\omega_{0\uab} &=& (\partial_j N + \eta N^i~ K_{ij})~ (n_\ua~ L^{j}_\ub -  n_\ub~ L^{j}_\ua )
 \nonumber  \\ 
& & - \nabla_{[j} N_{k]}~ L^{j}_\ua~ L^{k}_\ub + \eta~  n_{[\ua}~ 
\partial_t n_{\ub]} +  L_{j[\ua}~ \partial_t L^{j}_{\ub]}~.
\ee
The extrinsic curvature is
\be
K_{ij} &=& -\frac{1}{2N} \left( \partial_t g_{ij} - 
\nabla_i N_j - \nabla_j N_i\right)~.
\ee

From the point of view of the Lagrangian formalism, 
the momenta are just notation for expressions involving fields and their
derivatives
\be
p^{i\ua} &=& 2 \pi^{ij} L_{j}^\ua -
 \frac{1}{8} \eta~ \sqrt{\hat{g}}~ \bar{\psi} M \{ \Gamma^n, \Gamma^{\uab} \} 
\psi~ L_\ub^i \\
\pi^{ij} &=&- \frac{\eta}{\kappa^2}~ \sqrt{\hat{g}}~ (\hat{g}^{ij} \tr K - K^{ij})  \\
E^{\hat{A}} &=& \sqrt{g}~ \left( \frac{1}{\lambda^2} F^{0\hat{A}} + J^{0 \hat{A}} \right)  \\
\bar\chi &=&  \half \eta\sqrt{\hat{g}}~ \bar\psi~ \Gamma^n M \\
\chi &=&  \half \eta\sqrt{\hat{g}}~ \Gamma^n M \psi~.
\ee
The Poincar\'e constraints consist of three parts 
${\cal H}^\mu ={\cal H}^\mu_{I}+{\cal H}^\mu_{II}+{\cal H}^\mu_{III}$. 
In the pure gravity sector we have 
\be
  {\cal H}^\perp_I &=&  \sqrt{\hat{g}} 
\left( -\frac{1}{\kappa^2} {R} + \Lambda \right)
  - \frac{\eta\kappa^2}{\sqrt{\hat{g}}} \left( \tr \Pi^2 - \frac{1}{d-1} 
(\tr \Pi)^2 \right)  \\
  {\cal H}^i_I     &=& -\nabla_j (P^{j\ua} L^i_\ua)~, 
\ee
where the gravitational momenta have been shifted according to
\be
\Pi^{ij} =\pi^{ij}-\half \hat{g}^{ij}G \\
P^{j\ua} =p^{j\ua}-L^{i\ua} G~.
\ee
In the form field sector
\be
  {\cal H}^\perp_{II} &=&  
  \sqrt{\hat{g}} \left(- \frac{1}{2 \lambda^2 } F_{\hat{A}} F^{\hat{A}} - 
F_{\hat{A}}~ J^{\hat{A}}  \right)  \nonumber \\
& & + \frac{1}{\sqrt{\hat{g}}} \frac{\eta\lambda^2}{2} 
\left(E_{\hat{A}} - \sqrt{{g}} J^0_{~\hat{A}}\right) \left(E^{\hat{A}}
 - \sqrt{{g}} J^{0\hat{A}}\right)\\
  {\cal H}^i_{II}     &=&  F^i_{~\hat{A}} (E^{\hat{A}} - \sqrt{{g}} 
  J^{0\hat{A}})~,
\label{dyncurrent}
\ee
the different signs in front of the  fermionic 
dynamical (\ref{dyncurrent}) and background (\ref{backcurrent}) 
currents is not a surprise, 
as a contribution of the first
one has been used in the definition of the electric field $E_{\hat{A}}$.

In the fermion sector
\be
  {\cal H}^\perp_{III} &=&  \half \sqrt{\hat{g}} \left(
 - \bar\psi M \hat\Dslash \psi +  (\hat\Dslash \bar\psi)M \psi  - 2 \bar\psi  
Z_{\hat{A}}\Gamma^{\hat{A}} \psi
\right) \\
  {\cal H}^i_{III}     &=&   \half \eta\sqrt{\hat{g}}~ \left( \bar\psi~ \Gamma^n M  \hat{D}^i \psi - 
(\hat{D}^i \bar\psi)  \Gamma^n M \psi + 2  \bar\psi Z^i_{~\hat{A}} \Gamma^n 
\Gamma^{\hat{A}} \psi  \right)\label{backcurrent}
\ee
The action of the covariant derivative on spinors is, by definition
\be
\Dslash \psi &=& \Gamma^\mu( \partial_\mu +\frac{1}{4} \omega_{\mu\uab} \Gamma^\uab) \psi \\
(\Dslash \bar\psi) &=& (\partial_\mu\bar\psi - \frac{1}{4} \bar\psi 
\omega_{\mu\uab} \Gamma^\uab) \Gamma^\mu~,
\ee 
so that
\be
\tilde\nabla_\mu ( \bar\chi \Gamma^\mu \psi) = (\Dslash \bar\chi) 
\psi + \bar\chi \Dslash \psi~.
\ee

In addition to the Poincar\'e constraints there are also constraints that generate the 
gauge transformations $A \longrightarrow A + \dd B$ and gauge 
transformations on the frame bundle (local Lorentz transformations)
\be
  {\cal G}^{\hat{A}} &=&  \partial_i E^{i\hat{A}}\\
  {\cal J}^{\uab}    &=& p^{i[\ua} L_i^{\ub]} + \frac{1}{8} 
\eta~ \sqrt{\hat{g}}~ \bar{\psi} M \{ \Gamma^n, \Gamma^{\uab} \} \psi~.
\ee

\section{Clifford algebra} 

The formulae in this appendix are taken mostly from Ref.~\cite{Sohnius:1985qm}.

\subsection{Even dimensions}
\label{evencl}

Consider a metric with signature $\eta_{ab} = \{(-)^{d_-}, (+)^{d_+} \}$, 
such that $d = d_- + d_+ = 2 m$. 
The Clifford algebra is span by
\be
\{ \Gamma^i, \Gamma^j \} &=& 2 g^{ij}~.
\ee
All representations are unitarily equivalent, and the intertwining operators are 
\be
\Gamma^{\dagger}_{i} &=& {A} \Gamma_i {A}^{-1} \\
-\Gamma^{T}_{i} &=& {C}^{-1} \Gamma_i {C} \\
-\Gamma^{*}_{i} &=& {D}^{-1} \Gamma_i {D} \\
\Gamma^{*}_{i} &=& \tilde{D}^{-1} \Gamma_i \tilde{D}
\ee
where $D = C A^T$ and $\tilde{D} = \Gamma C A^T$. 
The chirality operator is
\be
\Gamma &=& \Gamma^1 \cdot \ldots \cdot \Gamma^d
\ee
We have the following phases
\be
 A &=& \alpha A^\dagger \\
 C &=& \tilde{\eta} C^T \\
 D D^{*} &=& \delta \\
 \tilde{D} \tilde{D}^{*} &=& \tilde{\delta} 
\ee
where  $|\alpha| = 1$, $\tilde\eta =\pm 1$, and $\delta^* =  \tilde{\delta}$. 
The Dirac conjugate is defined as $\bar{\psi} = \psi^\dagger {A}$, and the
charge conjugate as $\psi^c = {C}{A}^T\psi^*$, or  $\psi^c = \Gamma {C}{A}^T\psi^*$.
The chirality matrix satisfies
\be
\Gamma^\dagger &=& (-)^{m}~ A \Gamma A^{-1} \\
\Gamma^T &=& (-)^m~ C^{-1} \Gamma C~.
\ee

\subsection{Odd dimensions}

In this appendix we build a representation of an odd-dimensional $D=d+1$ Clifford algebra 
with $(\Gamma^n)^2=\eta$ starting from a given even dimensional $d=2m$ Clifford algebra.
We actually only need to construct the correct Clifford matrix $\Gamma^n$ 
\be
\Gamma^n &=& \sqrt{(-)^{m+d_-} \eta \hat{g}}~  \Gamma~.
\ee
In odd dimensions not all representations are equivalent. Instead, 
we only have the intertwining operators
\be
\Gamma^{\dagger}_{\ua} &=& (-)^{d_-}\eta~  \tilde{A} \Gamma_\ua \tilde{A}^{-1} \label{epsi} \\
\Gamma^{T}_{\ua} &=& (-)^{m}~ \tilde{C}^{-1} \Gamma_\ua \tilde{C} \\
\Gamma^{*}_{\ua} &=&  (-)^{m+d_-}\eta~ {\cal D}^{-1} \Gamma_\ua {\cal D}
\ee
where ${\cal D} =  \tilde{C} \tilde{A}^T$, not to be confused with (\ref{bigd}).
There are two inequivalent conjugacy classes: which 
representations belong to which depends on $m, d_-$ and $\eta$. 

We can represent the odd-dimensional intertwinors in terms of their even dimensional 
counter parts as
\be
(-)^{d_-} \eta = \left\{
\begin{array}{cll}
+1, &  \tilde{A} = A \\
-1, &  \tilde{A} =  A \Gamma^n
\end{array} \right. \label{AAtilde} \\
(-)^{m} = \left\{
\begin{array}{cll}
+1, &  \tilde{C} = \Gamma^n C \\
-1, &  \tilde{C} =  C
\end{array} \right. 
\ee
We sometimes abbreviate the sign $(-)^{d_-} \eta$ by $\varepsilon$. 
The operators  ${\cal D}$ can be expressed in
terms of boundary operators
\be
\begin{array}{|c|cc|} \hline
{\cal D} & (-)^m = 1 & (-)^m = -1 \\ \hline 
&& \\ 
(-)^{d_-}\eta~ =  1 & \tilde{D} & D \\ 
&& \\ 
(-)^{d_-}\eta~ =  -1 &\eta D & - \tilde{D} \\ 
&& \\ 
\hline
\end{array}
\ee
where $D = C A^T$ and $\tilde{D} = \Gamma^n D$.

\section{Fermion phase space}
\label{fermi}
\label{chiral}

Fermions can be decomposed in bulk and boundary components in essentially
two Lorentz invariant ways, namely by using chirality or reality conditions.
Here we consider only chirality conditions, with which
we refer to the eigenvalues $\pm 1$ of $\sqrt{\eta} \Gamma^n$. This is
essentially the only Clifford matrix whose 
eigenvalues we can consider without breaking Lorentz invariance explicitly. 

We have to divide our analysis in two cases depending on the
sign ${\eta}$ and details of the bulk metric. Defining 
$\sqrt{\eta} \Gamma^n~ \psi_\pm = \pm \psi_\pm$ we get 
\be
\bar{\psi}_\pm ~\sqrt{\eta} \Gamma^n = \pm \varepsilon \eta ~\bar{\psi}_\pm,
\ee
where $\varepsilon$ is the sign that appears in the relation of the 
Clifford matrices to their Hermitian conjugates (\ref{epsi}). This means that 
the the Dirac dual of a spinor of definite chirality is either of the same or the 
opposite chirality; as a consequence, the Lagrangian fields that correspond 
to the phase space coordinates will be different. It is useful to note that
\be
\varepsilon \eta = 
\left\{ 
\begin{array}{cc}
(-)^{d_-} & d+1 ~~ \mbox{odd} \\
\eta & d+1 ~~ \mbox{even}
\end{array}
\right.
\ee

{\bf Case I}. Assume $\varepsilon \eta = 1$.
The kinetic term separates into
\be
& & - \sqrt{\eta \hat{g}} ( \bar\psi_- M \dot\psi_- +  \dot{\bar\psi_+}  M \psi_+ )  
- \partial_t {G} \\
& & -\half \bar\psi_- \partial_t ( \sqrt{\eta \hat{g}} M)~  \psi_- 
 - \half \bar\psi_+ \partial_t ( \sqrt{\eta \hat{g}} M)~  \psi_+ \label{kinfermi}
\ee
where
\be
{G} &=&  -\half  \sqrt{\eta \hat{g}} ( \bar\psi_- M \psi_- +  
{\bar\psi_+}  M \psi_+ ) ~.
\ee
The fermionic phase space consists therefore of the symplectic pairs 
$(\varphi, \bar\pi)$ and $(\bar\varphi, \pi)$, where 
$\varphi = \psi_-$ and $\bar\varphi = \bar\psi_+$ and
\be
\bar\pi &=&  - \sqrt{\eta \hat{g}}~ \bar\psi_- M \\
\pi &=&  \sqrt{\eta \hat{g}}~ M \psi_+~.
\ee
The last two terms in (\ref{kinfermi}) produce a term
\be
\half \hat{g}^{ij} G ~ \partial_t {g}_{ij}
\ee
in the action, and therefore cause a shift in the gravitational momentum.

{\bf Case II}. Assume $\varepsilon \eta = -1$.
The kinetic term separates into
\be
 & & - \sqrt{\eta \hat{g}} ( \dot{\bar{\psi}}_- M \psi_+ +  {\bar\psi}_+  M \dot\psi_- ) 
- \partial_t {G}  \\
& & -\half  \bar\psi_- \partial_t ( \sqrt{\eta \hat{g}} M)~  \psi_+ 
 - \half  \bar\psi_+ \partial_t ( \sqrt{\eta \hat{g}} M)~  \psi_-  \label{kin2}
\ee
where
\be
{G} &=& - \half   \sqrt{\eta \hat{g}} ( \bar\psi_- M \psi_+ +  
{\bar\psi_+}  M \psi_- )~.
\ee
The configuration space is span by
$\varphi = \psi_-$ and $\bar\varphi = \bar\psi_-$, and the momenta are 
\be
\bar\pi &=&  - \sqrt{\eta \hat{g}}~ \bar\psi_+ M \\
\pi &=&  \sqrt{\eta \hat{g}}~ M \psi_+~.
\ee 
Notice that, due to (\ref{AAtilde}), the Dirac conjugates $\bar\varphi$ and $\bar\psi_-$ 
are formed differently: the former using the matrix $A$ and the latter with $\tilde{A}$. 
This will result in an extra $\Gamma^n$ everywhere, including the kinetic term, 
and the resulting extra sign hence cancels out.
The gravitational momenta are shifted as in Case I.

\end{document}